# Numerical Investigation of the Kinetics of Non-Equilibrium Phase Transitions in Silicon Induced by an Ultra-Short Laser Pulse

Dmitry S. Ivanov[1] and Tatiana E. Itina[1*]

[1] Université Jean Monnet Saint-Etienne, CNRS, Institut d'Optique Graduate School, Laboratoire Hubert Curien UMR 5516, F-42023, Saint-Etienne, France.

*tatiana.itina@univ-st-etienne.fr

## Abstract

Modern semiconductor applications demand precise laser processing at the nanometer scale, requiring a detailed understanding of phase transitions and structural modifications. Accurate control over laser-induced processes in semiconductors is essential for generating surface structures and modifying surface properties. In this study, we present a numerical investigation of non-equilibrium laser-induced phase transitions in silicon (Si) using a hybrid atomistic-continuum model. The model combines the strengths of Molecular Dynamics (MD) simulations for atomistic-scale descriptions of non-equilibrium phase transitions with a continuum approach to account for laser-generated free carriers. This framework captures the generation and diffusion of electron-hole pairs, thermal diffusion, and electron-phonon coupling during laser energy deposition. We apply the model to determine the melting depth as a function of fluence for a 100 fs laser pulse at 800 nm. The results show that the stand-alone continuum approach underestimates the melting threshold compared to the hybrid atomistic-continuum model, which incorporates the detailed kinetics of melting. Additionally, we explore the effect of crystal orientation on melting dynamics. Lastly, the MD model is used to identify the conditions leading to the amorphization of the Si surface. These findings provide valuable insights into experimental observations of Si surface structuring induced by ultrashort laser pulses.

## 1. Introduction

Silicon (Si) is one of the most important materials in modern technology, especially in microelectronics, optoelectronics, and solar cells. As the demand for faster, smaller, and more efficient devices continues to grow, being able to control material properties at the nanoscale is crucial. One of the keyways to achieve this is laser processing, which allows us to modify materials with high accuracy and on a nanometer scale [1,2]. In microelectronics, lasers are used for many purposes, including doping, annealing, and etching of semiconductors. These processes are essential for creating the tiny, complex patterns needed in integrated circuits. Ultrafast laser pulses, with their short duration, allow for extreme precision, minimizing damage to surrounding material. The ability to create localized phase changes, such as surface melting or amorphization, allows fine control of material properties, such as electrical conductivity or optical reflectivity [3,4]. Laser-induced amorphization of silicon is particularly useful for phase-change memory, a technology used in data storage, where switching between the crystalline and amorphous states can store information [5]. Understanding how these laser-induced phase transitions happen is critical for improving current and future microelectronics technologies [6].

Many experiments have been done to study how short and ultrashort laser pulses affect silicon. These studies have shown that ultrafast pulses can cause rapid melting of silicon, followed by the formation of an amorphous layer if the material's cooling rate is too high for recrystallization. Using ultrafast time-resolved techniques, scientists can observe how silicon melts and re-solidifies in real-time. The melting process depends on factors like the energy of the laser pulse (fluence), pulse duration, and crystal orientation. For example, at high fluences, *homogeneous melting* can occur, where the entire structure melts almost at once, driven by significant overheating exceeding crystal stability [7]. In contrast, *heterogeneous melting* starts at surfaces, defects, dislocations, or imperfections in the material, requiring lower temperatures and occurring unevenly across the sample. Pressure and temperature interplay key roles in these mechanisms—higher temperatures tend to favor homogeneous melting, while pressure, depending on the material, can suppress and enhance in. On the other hand, heterogeneous melting, depending on the presence of defects, is always limited in the speed of solid-liquid propagation, and can be overwhelmed by homogeneous mechanism of melting [8]. The experimental measurements, however, are also limited in what they can reveal about atomic-scale events, so we rely on simulations for complementary research of these processes.

To understand better the laser-induced phase transitions, several models were developed that simulate the complex laser-induced processes in solids, including semiconductors [9,10,11]. One of the common and most used methods is the Two-Temperature Model (TTM), which shows how the deposited energy is transferred between electrons and the atomic lattice after the laser pulse. However, this model alone can't capture important atomic-level details, such as how melting and amorphization happen. To solve this, researchers combine TTM with Molecular Dynamics (MD) simulations. In particular, because of the additional descriptor of the free carrier density $n$, for semiconductors this hybrid approach, known as the nTTM-MD model [], accounts for both large-scale energy transfers and small-scale atomistic description of the non-equilibrium phase transition. The nTTM-MD model is especially useful because it includes how free carriers (electron-hole pairs) generated by the laser affect the material, allowing us to study processes like energy transport, melting, and amorphization in detail.

One of the biggest challenges in modeling laser-induced phase transitions in silicon is accurately predicting the melting threshold and depth, especially at lower fluences. Traditional continuum models often underestimate the extent of melting because they don't fully account for atomic-scale processes. Our study, using the nTTM-MD model [12], shows a much better prediction of melting dynamics, particularly at low fluences. We explore the difference between heterogeneous and homogeneous melting and show how each mechanism leads to different outcomes in melting depth. We also study the effects of crystal orientation on melting. Our results show that crystal orientation significantly affects how silicon melts, with different orientations producing different melting patterns and rates.

Another important challenge is controlling the depth of amorphization. While laser-induced amorphization can be useful in some cases, such as memory storage, it can be a problem in microelectronics, where it might reduce electron mobility or change the material's electrical properties. Reducing the depth of amorphization without sacrificing the ability to make fine changes to the material's structure is a key goal. Our study focuses on refining the hybrid nTTM-MD model to better control melting and recrystallization while minimizing unwanted amorphization.

In this paper, we use a hybrid atomistic-continuum model (nTTM-MD) to study how free carrier temperature, lattice temperature, and free carrier density evolve during ultrafast laser excitation at a wavelength of 800 nm. We also investigate the relationship between melting depth and fluence for a 100 fs laser pulse, identifying the roles of heterogeneous and homogeneous melting. By comparing atomic configurations of silicon samples with <001> and <111> crystal orientations, we show how crystal structure affects the melting process. Finally, we discuss strategies for reducing amorphization depth.

The paper is organized as follows: Section 2 explains modeling details for both nTTM and the hybrid nTTM-MD model, parameters and numerical methods. Section 3 presents our simulation results, including free carrier dynamics, melting depth, and the effects of crystal orientation. In Section 4, we discuss how these findings relate to experimental results and applications in microelectronics. Final Section provides conclusions and future directions for the work.

## 2. Modeling details

The continuum part of the developed combined atomistic-continuum model is based on the approach proposed by van Driel [13], and further developed by Chen et al in Ref. [14]. The corresponding description of the effect of laser-generated free carriers is similar to a so-called Two Temperature model (TTM), which was formulated for metals by Anisimov et al in Ref. [15]. Unlike the TTM, however, where the conservation of energy is due to the balance between free carriers and phonons an addition equation on the carrier's density $N_e$ is required to account for multiphoton and free carrier's absorption process, impact ionization, Auge recombination and the carrier's flux (carrier's diffusion). In this way, the modified TTM for Si material describes the dynamics of free carrier's density as well is referred as nTTM. This model is based on the following system of equations [14]:

$$\begin{cases} \frac{dI(z,t)}{dz} = -\left(\alpha + \alpha_{FCA}\big(\varepsilon(N_e)\big)\right) I(z,t) - \beta I^2(z,t) \\[4pt] \frac{\partial N_e}{\partial t} = \frac{\alpha I(z,t)}{h\nu} + \frac{\beta I^2(z,t)}{2h\nu} + \delta N_e - \gamma N_e^3 - \nabla \cdot \vec{J}(N_e, T_e, E_g) \\[4pt] \frac{\partial U_e(N_e T_e, E_g)}{\partial t} = \left(\alpha + \alpha_{FCA}\big(\varepsilon(N_e)\big)\right) I(z,t) + \beta I^2(z,t) \\[4pt] \quad - \nabla \cdot \vec{W}(N_e, T_e, E_g) - \frac{C_{e\text{-}h}(N_e, T_e, E_g)}{\tau_e}(T_e - T_l) \\[4pt] C_l \frac{\partial T_l}{\partial t} = \vec{\nabla} \cdot \left(\kappa \vec{\nabla} T_l\right) + \frac{C_{e\text{-}h}(N_e, T_e, E_g)}{\tau_e}(T_e - T_l) \end{cases} \tag{1}$$

where $\alpha$, $\beta$, $\delta$ and $\gamma$ and are coefficients of correspondingly 1-, 2-photon absorption, impact ionization, and Auger recombination due to the irradiation of the intensity $I$, and $\alpha_{FCA}$ is free carriers absorption coefficient. The free carriers (electron-hole pairs) and phonons interact via the carrier-phonon coupling, which is defined here as a ratio of free carrier's heat capacity $C_{e\text{-}h}$ and the carrier-phonon energy relaxation time $\tau_e$. Noteworthy, the energy flux $W$ in Eq. (1) includes both the carrier's flux $J$ and heat diffusion defined via the conductivities of both electrons and holes.

The total energy of free carriers $U_e$ in Eq. (1) is, therefore, given by the sum of both the kinetic and potential energies, which comprise the energy gap $E_g$ (that varies during the laser-irradiated target evolution) as well [13]:

$$U_e = N_e E_g(N_e, T_e) + \frac{3}{2} N_e k_b T_e \left[ H_{1/2}^{3/2}(\eta_e) + H_{1/2}^{3/2}(\eta_h) \right] \tag{2}$$

where $H_j^i(\eta_c) = F^i(\eta_c)/F^j(\eta_c)$, and $F^j(\eta_c)$ is Fermi-Dirac integrals of order "$\underline{i}$" of the carriers with the reduced chemical potential $\eta_e = (E_F\text{-}E_c)/kT_e$ and $\eta_h = (E_V\text{-}E_F)/kT_e$ of correspondingly the electrons and holes with respect of their quasi Fermi level to the levels of conduction and valence bands correspondingly.

While the parameters entering the continuum part of the combined model can be varied and justified either by experimental measurements [13] or theoretical calculations [16,17], the strategy used in the development of the combined model for Si remains the same as the one used for MD-TTM for metals [18]. Namely, we replace the equation. in the system (1) for lattice temperature with MD integrations. As fot the interatomic potential $U(r_{ij}, T_e)$ for Si, either classical Stillinger-Webber parametrization [19] can be used, or one can include the dependences on the electronic temperature [20] and possibly on the electronic density [21].

$$m_i \frac{d\vec{r}_i}{dt^2} = -\sum_{j=1,N} \vec{\nabla} U(\vec{r}_{ij}, T_e) + \xi m_i \vec{V}_i^T \tag{3}$$

where $r_i$ is the position of "$\underline{i}$" atom with the mass $m_i$, $N$ is the number of atoms in the discretized cell, $V_i^T$ is thermal velocity of the atom "$i$", and the coefficient $\xi$ is determined from the energy conservation law [18]. Because of the additional equation for the carrier's density, we refer to this combined method as MD-nTTM [12 as schematically presented in Fig. 1.

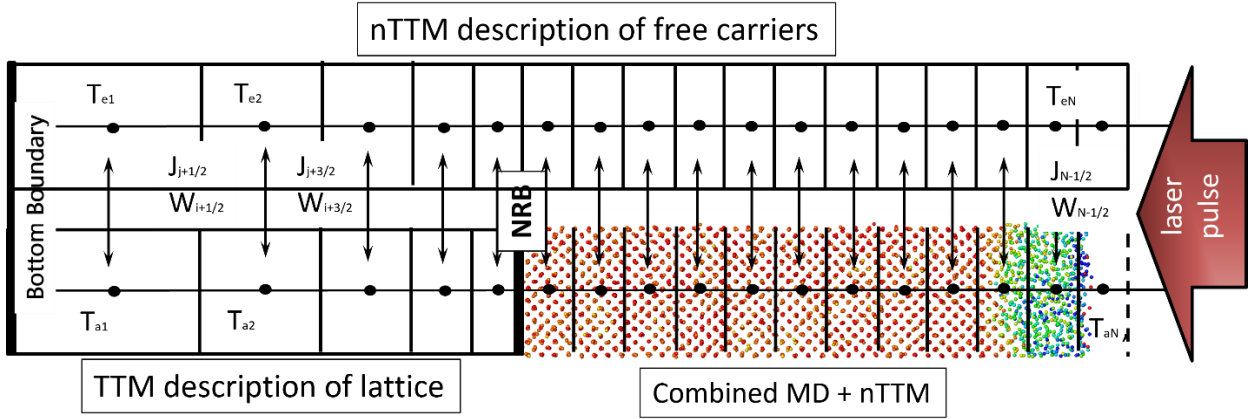

**Figure 1.**
Schematic representation of the combined MD-nTTM model for silicon (Si). The system of free carriers is described in the continuum (upper part), while the atomistic behavior is modeled using Molecular Dynamics (MD) (bottom part). This integrated approach enables an accurate description of non-equilibrium phase transitions. The free carrier system, modeled by nTTM, is divided into a number of cells, with each cell corresponding to the same volume in the MD portion. The nTTM framework captures the processes of laser light absorption, free carrier generation, and their subsequent diffusion and thermal diffusion. By synchronizing the time steps of nTTM and MD, where $\Delta t_{MD} = n\Delta t_{nTTM}$, the energy transfer from hot carriers to the corresponding MD cell is equally distributed over the atoms constituting that MD cell. For thicker targets, the application of Non-Reflective Boundary (NRB) conditions [22] allows one to limit the MD application to the top of the irradiated target only, leaving its bottom part of the computational cell described completely via nTTM model [18,23].

It is worth of mentioning that unlike a similar model for Si published in Ref. [24], where non-degenerate model based on the Boltzmann distribution of the electron-hole pairs was assumed (indicating negligible excitation), the complete description of the laser-generated free carriers dynamics via Fermi-Dirac functions is applied in the presented model. This important modification results in the better applicability of the MD-nTTM model for investigation of laser melting mechanism of Si at the fluences up to the ablation threshold and above.

To solve the proposed system (Eq. 1), the finite difference grid mesh is used, as sketched in Fig. 1. The sample is divided into cells according to the scheme, and the thermodynamic parameters are calculated in each cell. The spatial derivatives of $n$, $T_e$, $T_a$, $J$, $W$, and $E_g$ at the interior points are approximated with the central difference, and those at the boundary are evaluated with the first-order approximation. Given by the Neuman stability criterium, the explicit scheme of the finite differences method, however, results in an extremely shallow time step ($\sim 10^{-25} \div 10^{-22}$ s), depending on spatial discretization, provided by the required resolution, which typically in the range of 1-10 nm due to modern technological requirements. As the MD time step is normally as small as $\sim 10^{-15}$ s, the combined MD-nTTM model in this case becomes either very time-consuming or poor on resolution [25]. To overcome this obstacle, a more advanced implicit algorithm based on the Crank-Nicolson scheme [26] was utilized that allowed us to raise the nTTM time step by few orders of magnitude reaching the range of $10^{-17} \div 10^{-16}$ s [27], making therefore the suggested MD-nTTM approach an efficient numerical tool to study strongly non-equilibrium laser-induced phase transitions processes in Si. The complexity of implementation of Crank-Nicolson scheme, however, is frequently limiting the application of nTTM model, Eq. (1) to 1D only. Additionally, the implicit scheme of the Finite Differences (FD) method can be barely parallelized for multiprocessing regime, indicating the limited material volume for numerical processing as well. While this problem does not concern us in the present study, the necessity of implementation of Crank-Nicolson scheme in 3D and in multiprocessing regime for investigation of the Si surface functionalization/nanostructuring mechanisms becomes obvious. Likely, the elegant modification of Crank-Nicolson scheme in 3D was recently suggested in Ref. [28] and will be utilized in our future studies.

## 3. Results and Discussions

### 3.1 Free Carrier and Lattice Dynamics

To clearly demonstrate the applicability and the accuracy of the nTTM model implementing Crank-Nicolson algorithm, we perform the initial modeling for the ultra-short laser pulse with temporal width of 100fs at the incident fluence of 0.2 J/cm$^2$ interacting with 2000nm free standing Si film. To describe the reflectivity function and the absorption by free carriers we used Drude model as it was suggested in Ref. [16] assuming the wavelength of 800nm. In this case we use laser intensity is given by Gaussian distribution as follows:

$$I_0(t) = F\left(1 - R(\varepsilon)\right)\frac{1}{\tau}\sqrt{\frac{\sigma}{\pi}}\,exp\left(-\sigma\frac{(t-t_0)^2}{\tau^2}\right) \qquad (4)$$

where $F$ is the average incident fluence, $\tau$ is pulse duration, and $R$ is the reflectivity function that can depend on the level of excitation via the electronic temperature, $\sigma = 4ln2$, the pulse is centered at $t_0 = 2.5\tau$, and $\varepsilon$ is the complex dielectric function, defined as:

$$\varepsilon(\omega_L) = \varepsilon_r(\omega_L) - \frac{N_e e^2}{\varepsilon_0 \omega_L^2}\left[\frac{1}{m^*_{e,cond}\left(1+i\frac{v_e}{\omega_L}\right)} + \frac{1}{m^*_{h,cond}\left(1+i\frac{v_h}{\omega_L}\right)}\right] \qquad (5)$$

where $\varepsilon_r$ is the intrinsic dielectric constant, $\omega_L$ is the laser frequency, and $m^*_{c,cond}$ is the conductivity effective mass defined to reproduce electrical conductivity and susceptibility [29]. The Drude collision frequency of electrons and holes, respectively, is denoted as $v_c$. By supposing that the collision frequencies of electrons and holes are equal and by introducing a common optical effective mass $1/m^*_{opt} = 1/m^*_e + 1/m^*_h$, Eq. 5 gives an ordinary Drude model for dielectric function [30], where complex refractive index $n^* = \varepsilon^{1/2}$ and the reflectivity function R and free carriers' absorption are given correspondingly:

$$R = \frac{[n^*-1]^2}{[n^*+1]^2} \qquad \text{and} \qquad \alpha_{FCA} = \frac{Im(n^*)\omega_L}{c_0} \qquad (6)$$

where $c_0$ is the speed of light in vacuum.

Fig. 2 shows the calculated electron temperature, lattice temperature, and free carriers' density dynamics obtained by using the nTTM model together with equations (4-6).

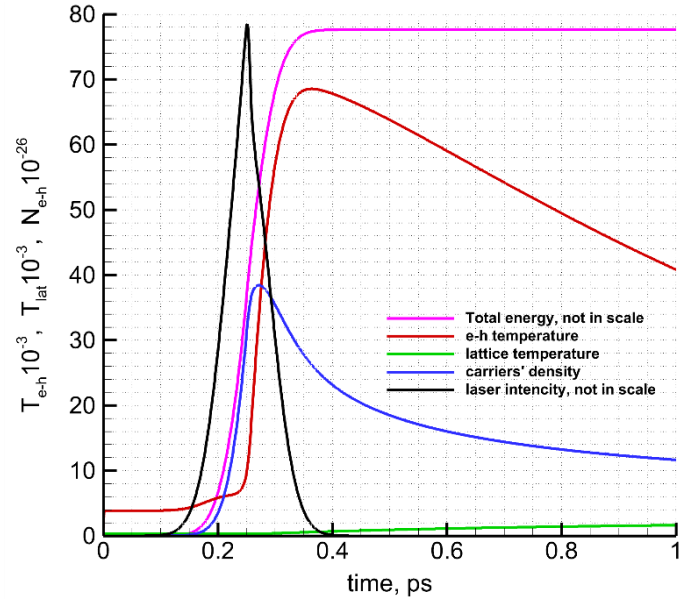

**Figure 2.** The dynamics of free carrier temperature, lattice temperature, and free carrier density are illustrated for the excitation of a 2000 nm free-standing silicon (Si) film using a 100 fs laser pulse with a wavelength of 800 nm at an incident fluence of 0.2 J/cm². Please note that the total energy conservation and the intensity of the laser irradiation are not to scale.

The system was spatially discretized into 400 cells (cell size is 5nm), which gives a high enough precision on one hand and the accuracy of the algorithm implementation is guaranteed by the total energy conservation (not in scale) on the other hand. The initial jump of the electronic temperature is explained by a negligible density of free carriers at the beginning of the pulse resulting in their extremely low heat capacity. The following deviation of the electronic temperature from what is normally observed in traditional TTM is due to the dynamically changing chemical potential and the energy gap as well. Then, the calculated lattice temperature and free carriers' dynamics, referred to the shape of the pulse intensity (not in scale) are reasonable.

### 3.2 Melting Depth Dependency on Laser Fluence

The relevant simplicity and efficiency of the nTTM numerical method, therefore, looks frequently attractive for the interpretation of many experimental findings [31]. The standing alone nTTM model, however, does not include the kinetics of melting mechanism of Si, and does not account on the peculiarities of it crystal structure under extreme conditions realized in the material upon ultrashort laser pulse irradiation. For that purpose, we perform MD-nTTM study of the melting depth in Si due to 100fs pulse at 800nm for a range fluences. The initial sample with <001> crystal orientation and dimensions of 5.5 x 5.5 x 1000 nm was prepared and equilibrated at 300K.

The sample thickness of 1000 nm was chosen based on the efficiency of the calculations and to ensure the thickness effect on the result. In fact, we obtain the same melting threshold for 2000nm sample as well. Finally, we run the MD-nTTM simulations for the range of fluences beginning with the melting threshold up to the ablation and above: $0.3 - 1.5$ J/cm$^2$. The results obtained with MD-nTTM model are compared with simulations given by nTTM for the same conditions and shown in Fig.3 (a).

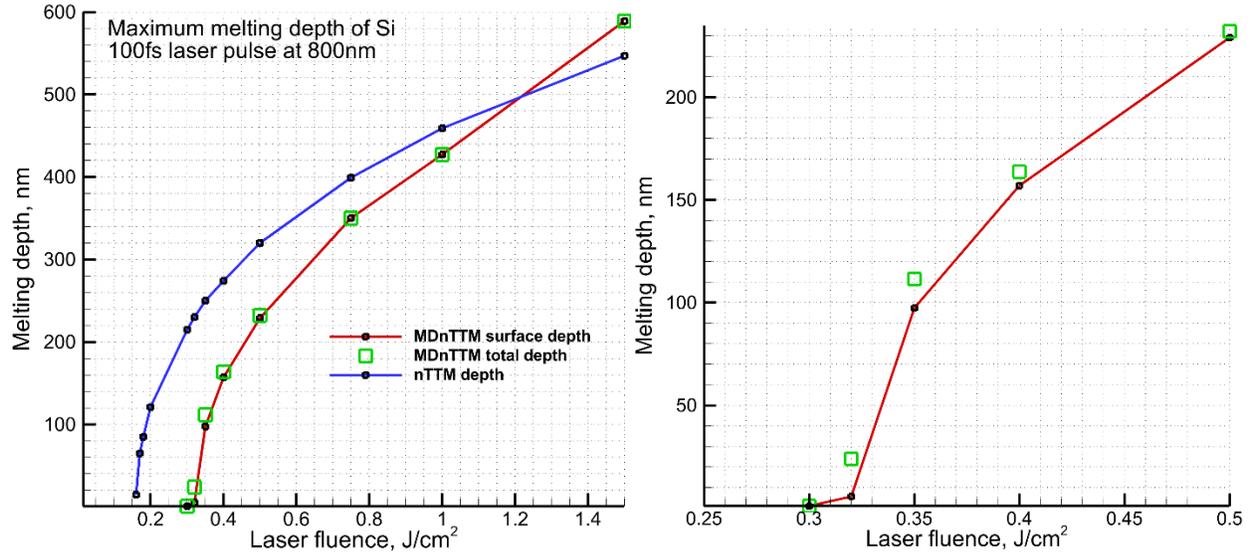

**Figure 3.** (a) The depth of melting calculated for a 100 fs laser pulse as a function of incident fluence at a wavelength of 800 nm is presented, as determined by the MD-nTTM and nTTM models. This analysis distinguishes between the depth of presurface melting due to the heterogeneous mechanism and the depth associated with the homogeneous melting mechanism. (b) A zoomed-in view of the melting depth as given by the MD-nTTM model at low fluences is also shown.

Interestingly, the melting depth given by the MD-nTTM model reveals quite a complex behavior as a function of laser fluence. First, the calculated melting threshold is equal to 0.3J/cm$^2$ in agreement with number of the previous experimental measurements obtained under similar conditions [32,33,34]. Second, the resulting melting depth at low fluence regime is clearly attributed by both heterogeneous and homogeneous melting mechanisms. This can be seen in Fig. 4, where the final atomic configuration reveals several melted layers, including the surface melting. Because of that, Fig. 3 (a) shows both the maximum melting depth due to heterogeneous phase transition mechanism and the green squares indicate the accumulated depth of melting due to homogeneous mechanism as well, Fig 3 (b). However, as laser fluence increases, the difference between both measurements disappears. This can be explained by the fact that under strong enough laser irradiation, the relatively slow heterogeneous melting mechanism contributes only slightly to the final state as compared to homogeneous melting mechanism that a massive character. And third, at the fluence above 0.7 J/cm$^2$ the melting depth starts scaling rather linearly, due to spallation mechanism of the target damage. This closely corresponds to the experimental measurements of Bonse group [35].

On the other hand, the melting depth obtained with nTTM model under the same conditions reveals much larger values (100 – 200 nm) at low fluences and even lower at high energies. The melting threshold is found at the fluence of 0.16 J/cm$^2$, which completely reproduces the result obtained in work by Ramer [16]. Such a large difference can be attributed to the absence of melting kinetics in the nTTM model, which is however naturally included in MD method. Also, the effect of pressure and the latent heat of melting are missing in nTTM. This results in an underestimation of the melting threshold provided by the continuum methods such as nTTM. At high fluence regime, however, the melting depth given by MD-nTTM approach becomes even higher. This is the result of the onset of the spallation regime of the material removal, which is also absent from the description due to nTTM stand alone. Here we refer to the spallation mechanism of the ablation process, when the material undergoes to the mechanical damage due to extraction of large chunks and even presurface layers.

### 3.2 Melting Kinetics and the Role of Crystal Orientation

The next part of our study is devoted to the effect of crystal orientation on the melting mechanism and depth of melting. There are many experimental measurements indicate the smaller thresholds of melting, ablation, and amorphization for Si crystal with <111> direction [36]. For that purpose, we prepared two samples of the same dimensions, but with two different crystal orientations <001> and <111>. Then, both samples were irradiated with 100 fs laser pulse at 800 nm at the same fluence 0.3 J/cm$^2$, that was above identified as the melting threshold. The modeling results can be seen in Fig. 4 revealing the melting dynamics of both samples as a sequence of the atomic configurations taken at the same times.

Despite a similar melting dynamics for both samples, the final result obtained with MD-nTTM model reveals that the sample with <001> orientation contained of about 25% more the material in liquid sated than the one with <111> orientation. This result is was not noticed in the above experimental measurements, but can be partially theoretically explained. Namely, as it is well-known, fast heating of the irradiated solid target due to ultrashort laser pulse induces a fast increase in the internal pressure as well [37]. The subsequent relaxation of the laser-induced stresses inside the material, however, can vary depending on the irradiation condition and, especially, on the crystal structure. Thus, it was shown in Ref. [12] that the effect of pressure on melting kinetics of Si is quite different from that of metals. For metals exhibiting fcc crystal structures, the increase of pressure leads to the reinforcement of crystal stability against the melting process. And phase transition itself results in an increase of material volume by 3-5%, whereas in case of semiconductor materials, exhibiting open diamond crystal structure we have an opposite effect. The increase in pressure in the solid phase of Si weakens its stability, and the material loses its volume upon the phase transition by 7%.

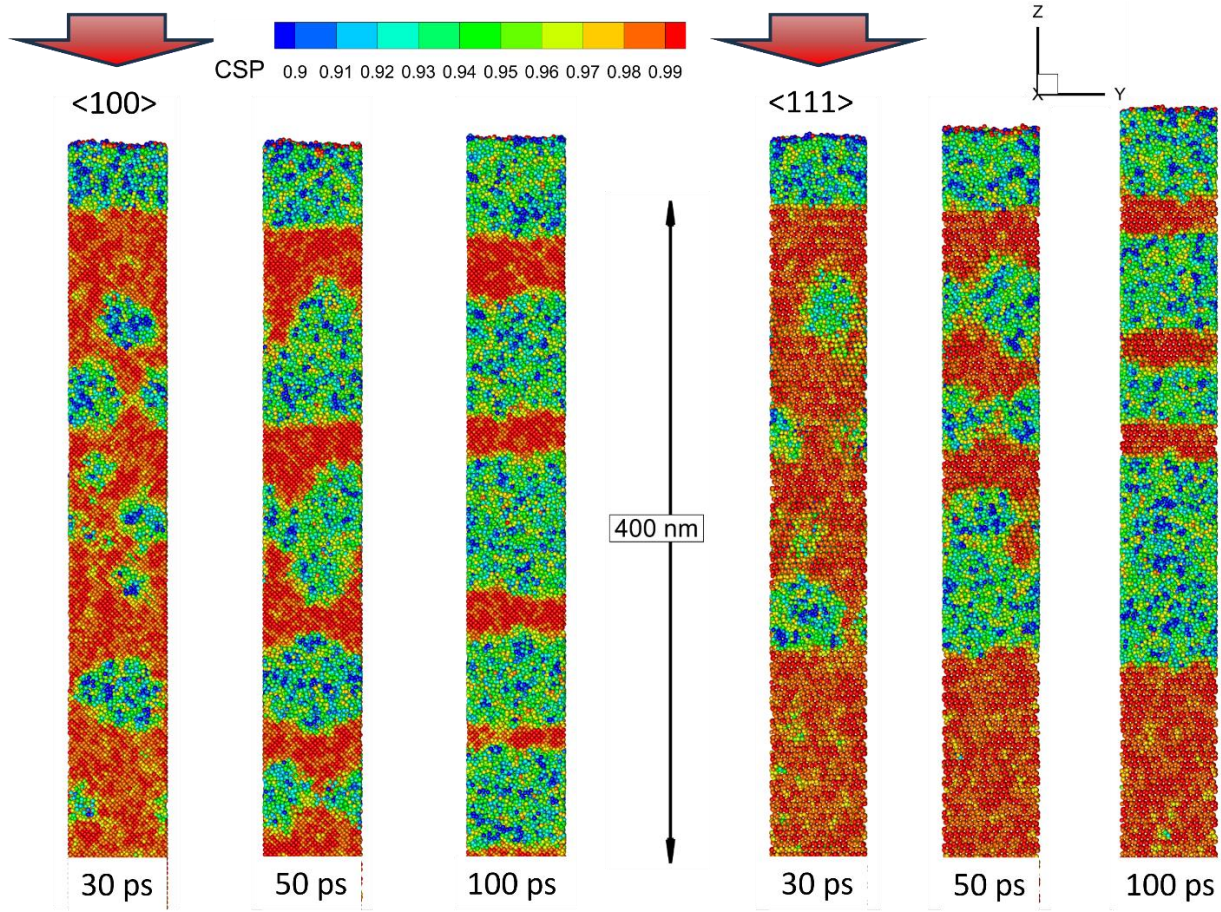

**Figure 4.** The sequence of atomic configurations illustrates the differences in the laser-induced melting process for Si samples with <001> (left) and <111> (right) crystal orientations. Both samples were irradiated with a 100 fs pulse at 800 nm, using the same incident fluence of 0.3 J/cm². Snapshots are presented at identical time intervals for both cases, with atoms color-coded according to the Central Symmetry Parameter (CSP), enabling the distinction between solid (CSP > 0.97) and liquid (CSP < 0.97) phases.

Initially rising internal stresses in Si due to ultrashort laser heating immediately relaxes upon the phase transition so that due to negative value of the volume of melting the internal pressure becomes negative as well [38]. In our simulations, as can be seen in Fig.4, the final configuration of the <111> crystal is expanded a bit more than the one with <001> orientation. This result can be due to a higher value of the speed of unloading pressure wave <111> Si. To verify this, we performed two identical MD simulations, in which the densification of atomic system by 5% was created at the edge of both samples. Following the pressure wave propagation it was found that the speed of the pressure wave is 8.1 nm/ps in <001> direction, whereas it is 8.74 nm/ps in <111>, that is by 8% greater. As a result, a faster relaxation of the <111> crystal leaves it in a preferable condition for supporting the solid phase. On the other hand, during the simulation process we used the same Drude-Lorenz model with the same parameterization as in Ref. [16]. We note, however, that there are studies considering the optical properties of Si to be dependent on the crystal orientation [39]. Therefore, the difference in the absorption process, depending on the crystal orientation, can play a more important role and lead to a smaller melting threshold for the Si crystal with <111> direction.

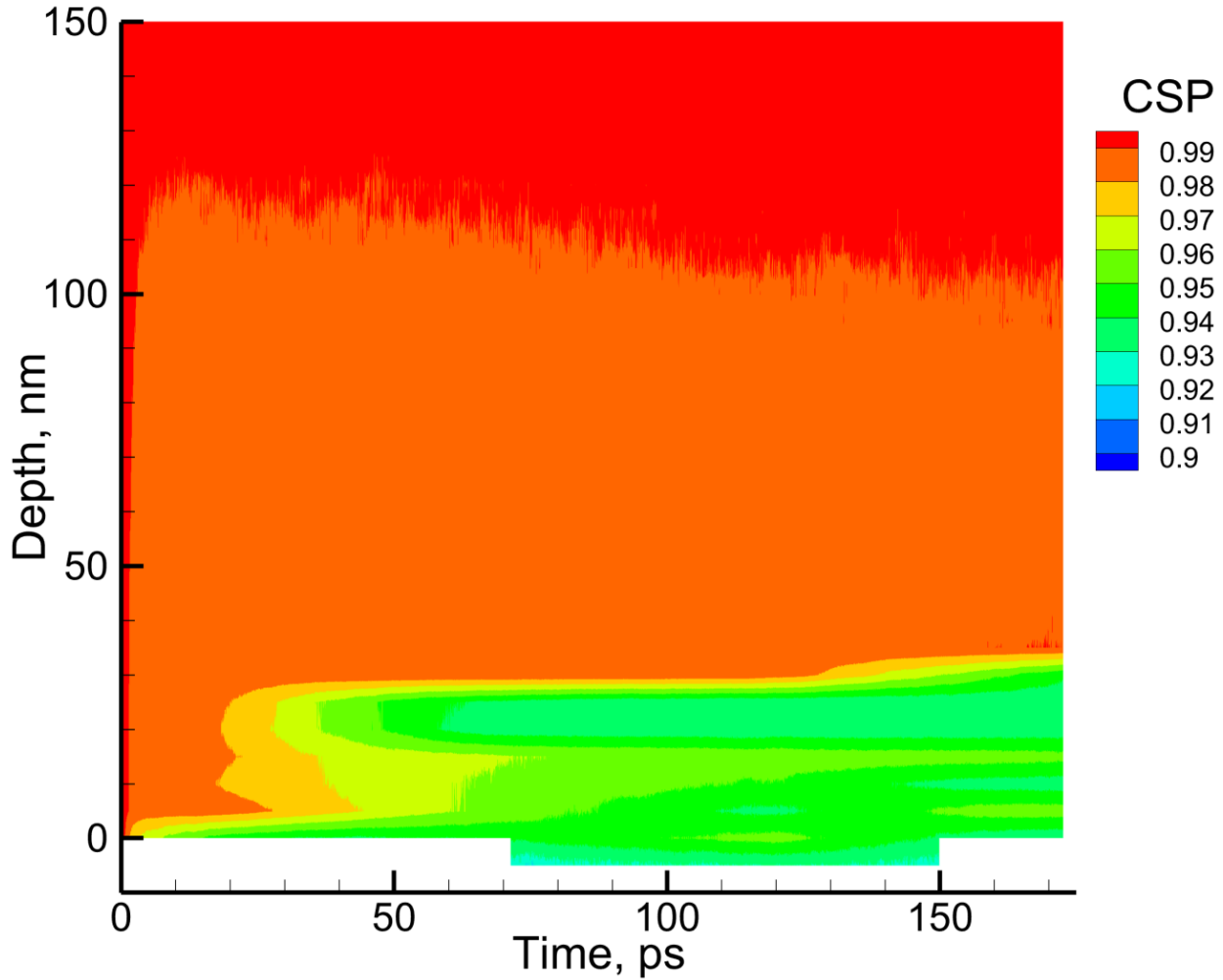

**Figure 5.** The dynamics of the CSP parameter for the irradiated Si target is illustrated as a contour plot for a 100 fs pulse duration at a wavelength of 800 nm and an incident fluence corresponding to the melting threshold. The CSP parameter is represented by color, and based on its empirically determined threshold value of 0.97, it reveals the melting process as a combination of both homogeneous and heterogeneous mechanisms.

To provide a more quantitative view of the melting process, we present the melting dynamics using the crystal structure parameter (CSP) in the contour plot shown in Fig. 5. This allows us to distinguish and quantify the contributions of two melting mechanisms: heterogeneous and homogeneous. Heterogeneous melting, a classical process, begins at the free surface and progresses into the bulk, though it is limited by the propagation speed. In contrast, homogeneous melting occurs when the material is overheated (~25%) above the melting point, making surface-driven melting impossible and causing rapid bulk melting. In Fig. 5, the yellow and orange regions likely represent the experimentally measured melting depth, consistent with Figure 4. However, fluctuations are observed before 20 ps, along with greenish areas that persist until the end of the simulation, and thin yellow-green regions appear around 50 ps below the main melting depth.

### 3.3 Laser-induced amorphization of Si

The amorphous phase of Si in the presurface region is frequently registered upon the ultrashort laser excitation [32] in the experiments and technologically attracts a high interest due to its unique properties [5]. The electrical and optical properties of the amorphous Si layer depend on its thickness, which in turn is a function of the laser irradiation parameters. So, a better understanding the mechanism of the amorphous layer formation is of a particular interest nowadays, particularly in microelectronics, but also to generate the well reproducible structures with specific parameters under controllable condition.

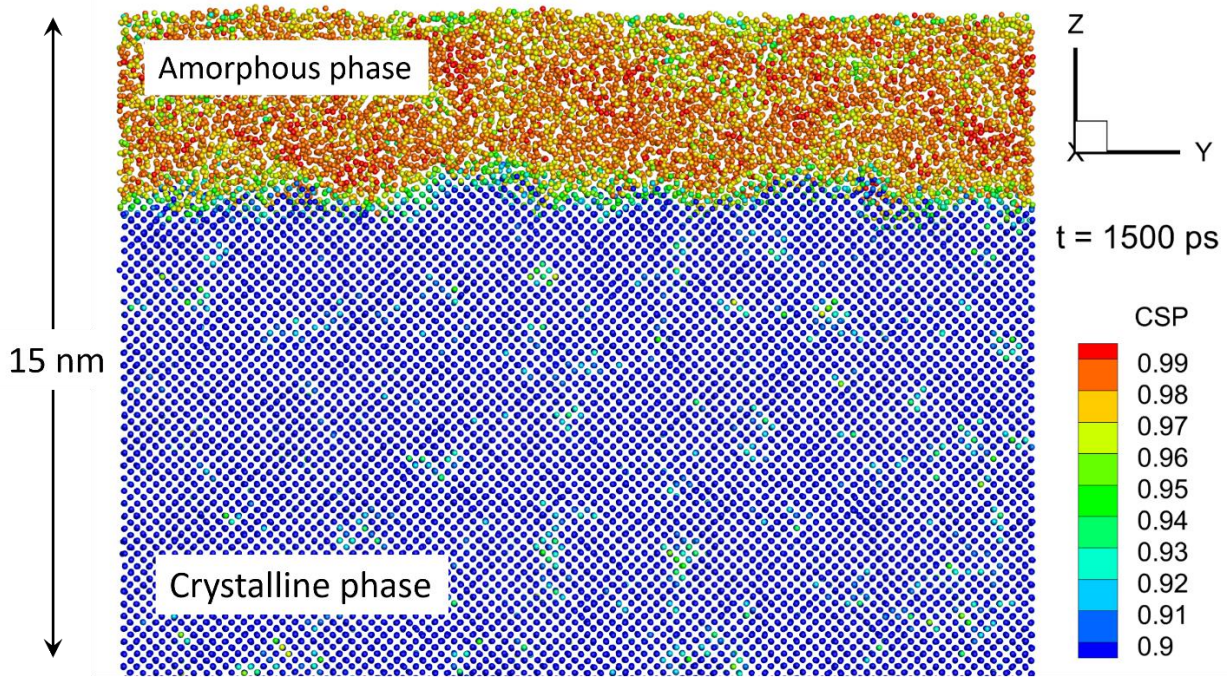

**Figure 6.** The presurface amorphous state of Si was achieved through a rapid cooling process of the melted region. Atoms are color-coded by the Central Symmetry Parameter (CSP), indicating that amorphization occurred due to the prolonged immobility of the solid-liquid interface, even after the sample temperature dropped to 1000 K, well below the melting point of 1687 K.

Finally, we identify the necessary cooling conditions in order to achieve the amorphous state of Si and relate the results to the optimal laser-Si interaction regime in experiments [32]. For that purpose, we prepare the initial sample at a temperature 25% above the equilibrium melting point and let it melt heterogeneously from the surface. Then, we try several cooling rates observing the process of recrystallization. Upon the achievement of an amorphous state, we relate the obtained cooling rate to the corresponding laser irradiation condition with the help of nTTM model. Since CSP does not distinguish between liquid and amorphous state, the amorphous state was registered id the solid-liquid interface propagation due to recrystallization remaining immovable for a long time enough. The resulting atomic configuration, where amorphous state was registered is shown in Fig. 6.

The cooling rate required for the formation of an amorphous state given by MD simulations was measured to be roughly 30K/ps. When trying to perform the nTTM simulations to match the above cooling rate conditions, however, the fastest cooling rate was found near the melting threshold and was measured at the level of 2.5 K/ps only, which is a way lower than the required one. The reason of such a large discrepancy could be again in inapplicability of the continuum approaches in the estimation of the recrystallization kinetics and also in the choice of the interatomic potential. We note that we chose the standard form of the parametrization of the Stillinger-Webber potential in this simulation [40], which describes well the thermodynamical properties of Si material. This gives us the possibility of a direct comparison of the modeling results with the experimental measurements. The modified Stillinger-Webber interatomic potential [41], previously use for the reproduction of the amorphous state, was not found suitable for the reproduction the melting point, which deteriorates the interpretation of the experimental results.

## Summary and Conclusions

This study presents a powerful and validated modeling approach to ultra-short laser interactions with silicon (Si), utilizing a hybrid method that combines the Two-Temperature Model (TTM) with Molecular Dynamics (MD), referred to as the nTTM-MD model. A notable advancement in our work is the enhancement of the Drude term in the laser interaction mechanism, facilitating a more precise description of free carrier generation and energy transfer during laser irradiation. This improvement allows for better predictions of material responses, particularly regarding phase transitions such as melting and amorphization.

Our findings yield several critical physical insights. We observe significant differences in melting behavior based on crystal orientation: samples oriented along the <001> direction exhibit more uniform melting, while those with <111> orientation display complex and uneven melting patterns. By analyzing the dynamics of free carrier temperature, lattice temperature, and free carrier density, we emphasize the essential role of laser-induced free carriers in driving phase transitions. This interaction is vital for accurately predicting the melting depth as a function of fluence, revealing distinct regions characterized by heterogeneous and homogeneous melting mechanisms.

A key result of this work is the identification of the conditions necessary for laser-induced amorphization of Si. We demonstrate that a rapid cooling process results in the formation of a presurface amorphous state, where recrystallization is hindered by the slow movement of the solid-liquid interface. This behavior holds particular significance for applications in phase-change memory and microelectronics, where precise control of amorphization depth is critical for optimal device performance.

The novelty of our work lies in the enhanced modeling accuracy, particularly at low fluences, and the clear distinction between heterogeneous and homogeneous melting mechanisms.

By effectively capturing both large-scale energy transport and atomistic-level phase transitions, our model bridges the gap between continuum and atomistic simulations, offering more reliable predictions for laser-material interactions.

Finally, our study opens several promising directions for future research. Further refinements to the model could involve incorporating multi-pulse irradiation effects and complex laser waveforms, which are often encountered in industrial applications. Additionally, extending the model to other semiconductor materials or integrating it with experimental time-resolved measurements could help validate and enhance its predictive capabilities.

## Author Contributions

DSI has developed the model, performed the calculations and analysis and has written the main part of this manuscript. TEI has conceived the ideas, models, worked on the problem analysis, edited the manuscript, and obtained funds for the project.

## Conflicts of interest

The author declares no conflicts of interests.

## Acknowledgements

The authors gratefully acknowledge the project ANR-23-CE08-0029 "Laser-forming of ultrathin amorphized Si layer for microelectronics applications" (LAMORSIM) for financial support. Computer support was provided by CINES of France under the project AD010814604R1